\documentclass[conference]{IEEEtran}
\IEEEoverridecommandlockouts

\usepackage{cite}

\usepackage{amsmath,amssymb,amsfonts}
\usepackage{subfigure}
\usepackage{algorithmic}
\usepackage{graphicx}
\usepackage{textcomp}
\usepackage[font=small]{caption}
\usepackage{dblfloatfix}
\usepackage{xcolor}
\usepackage{pdfpages}
\def\BibTeX{{\rm B\kern-.05em{\sc i\kern-.025em b}\kern-.08em
    T\kern-.1667em\lower.7ex\hbox{E}\kern-.125emX}}
\begin{document}

\title{Deformable-Detection Transformer for Microbubble Localization in Ultrasound Localization 
Microscopy\\
}

\author{\IEEEauthorblockN{Sepideh K. Gharamaleki}
\IEEEauthorblockA{\textit{Department of ECE} \\
\textit{Concordia University}\\
Montreal, Canada \\
sepideh.khakzadgharamaleki@mail.concordia.ca}
\and
\IEEEauthorblockN{Brandon Helfield}
\IEEEauthorblockA{\textit{Department of Physics and Biology,} \\
\textit{Concordia University}\\
Montreal, Canada \\
brandon.helfield@concordia.ca}
\and
\IEEEauthorblockN{Hassan Rivaz}
\IEEEauthorblockA{\textit{Department of ECE} \\
\textit{Concordia University}\\
Montreal, Canada \\
hrivaz@ece.concordia.ca}

}

\maketitle

\begin{abstract}
To overcome the half a wavelength resolution limitations of ultrasound imaging, microbubbles (MBs) have been utilized widely in the field. Conventional MB localization methods are limited whether by exhustive parameter tuning or considering a fixed Point Spread Function (PSF) for MBs. This questions their adaptability to different imaging settings or depths. As a result, development of methods that don't rely on manually adjusted parameters is crucial. Previously, we used a transformer-based approach i.e. DEtection TRansformer (DETR) \cite{Carion_DETR_ECCV2020,Gharamaleki_IUS2022} to address the above mentioned issues. However, DETR suffers from long training times and lower precision for smaller objects. In this paper, we propose the application of DEformable DETR (DE-DETR) \cite{Xie_DEDETR_CVPR2021} for MB localization to mitigate DETR's above mentioned challenges. As opposed to DETR, where attention is casted upon all grid pixels, DE-DETR utilizes a multi-scale deformable attention to distribute attention within a limited budget.
To evaluate the proposed strategy, pre-trained DE-DETR was fine-tuned on a subset of the dataset provided by the IEEE IUS Ultra-SR challenge organizers using transfer learning principles and subsequently we tested the network on the rest of the dataset, excluding the highly correlated frames. The results manifest an improvement both in precision and recall and the final super-resolution maps compared to DETR.

\end{abstract}


\section{Introduction}
Inspired by the myriad of approaches in microscopy to overcome diffraction limitations, such as Single-Molecule Localization Microscopy (SMLM) \cite{Lelek_SMLM_Nature2021} and photoactivated localization microscopy (PALM) \cite{Betzig_Science2006,Hess_Biojournal2006}, Ultrasound Localization Microscopy (ULM) has been developed to facilitate the visualization of microvasculture using ultrasound \cite{Errico_Nature2015}. 
ULM is a non-invasive method developed based on the precise tracking and precision of microbubbles (MBs) \cite{Christensen_UMB2019}. 
MBs, characterized by their similar size to red blood cells, have a high scattering coefficient and create harmonic and sub-harmonic frequencies which make them ideal for imaging small vessels with higher precision than conventional ultrasound.
The resonance of oscillating MBs within blood vessels increases the intensity due to an impedance mismatch between blood and gas \cite{BH_Frontiers2022}.

 Despite ULM surpassing the diffraction limit of conventional ultrasound, MB localization methods have difficulty pinpointing MBs that are not adequately isolated. Since low concentrations of MBs are used to ensure isolation between MBs, this results in a trade-off between image resolution and acquisition time. A longer acquisition time can prolong data collection and adversely affect imaging through motion artifacts. 
 
 Reliable MB localization is an essential step for super-resolution ultrasound, since inaccurate measurements propagate through the ULM pipeline and degrade the final ULM image's quality. Hence, in order to attain higher resolution and more accurate super-resolution maps, devising an adequate localization method is a key concept.
 
 The effects of MB signal cross-talks in dense concentrations of MBs have been the subject of many research papers such as \cite{Christensen_TUFFC2019} and \cite{Tanter_ScientificRep2019}. 
Further complications in the localization of the centroids of MBs arises, when fixed Point Spread Functions (PSFs) are utilized.
As a result of variation in space, phase aberration, and attenuation, uncertainty in PSF estimation leads to further vulnerability in the final calculated MB locations \cite{Christensen_TUFFC2017}.
 To ensure MB signal overlapping is prevented and to increase the localization accuracy of the MBs, diluted concentrations of MBs were used in \cite{Viessmann_PhysMedBio2013} and \cite{Reilly_MedPhys2013}.
  \begin{figure}[t]
\begin{minipage}[b]{1\linewidth}
\centering
\includegraphics[width=\textwidth]{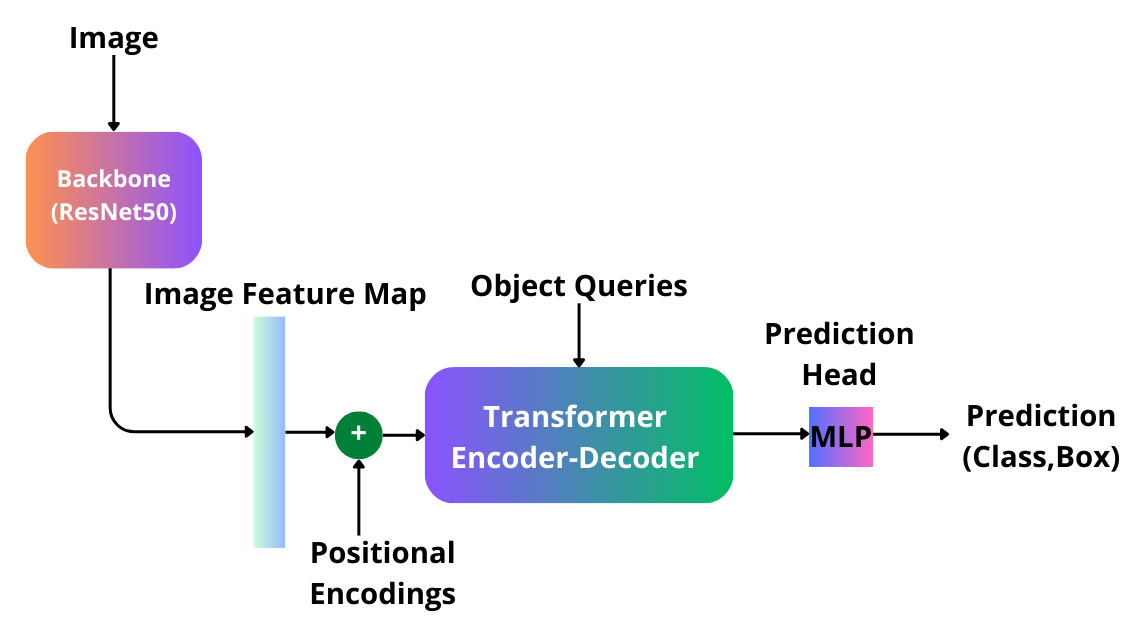}
\caption{DETR's architecture \cite{Carion_DETR_ECCV2020}.}
\label{fig:DETR}
\end{minipage}
\end{figure}
\begin{figure*}[b]
\begin{minipage}[tb]{1\linewidth}
\centering
\includegraphics[width=\textwidth,height=220pt]{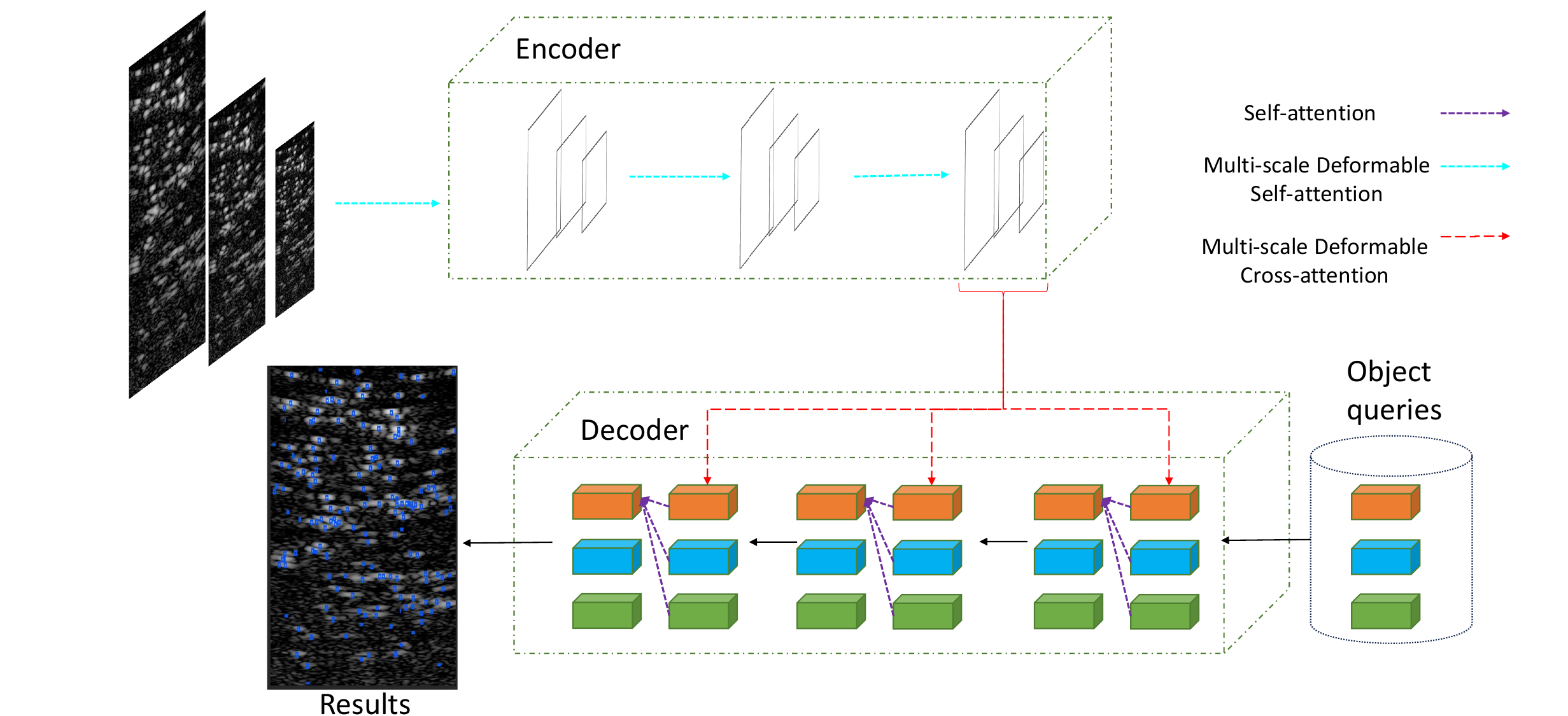}
\caption{DE-DETR's architecture (adapted from \cite{Xie_DEDETR_CVPR2021}).}
\label{fig:DEDETR}
\end{minipage}
\end{figure*}
In order to extract MB positions with high precision, signal processing algorithms were developed. In \cite{Ashik_IUS2022} correlation and amplitude-weighted center finding were utilized. In \cite{MLE_TUFFC2020}, maximum likelihood estimation was used to drive a statistical model for MB counts. Bayesian spectral estimation \cite{Sboros_IUS2018}, has been utilized for frequency estimation of non-linear MB signals, as well. While these methods required extensive parameter tunings, \cite{MLE_TUFFC2020} and \cite{Sboros_IUS2018} also performed poorly with dense concentrations of MBs. 
MB signal separation at high MB concentrations was also attempted using Fourier-based filtering \cite{Song_ScientificRep2020}.

Even though ULM imaging quality was improved, it did not perform as well in scenarios with higher MB counts or complicated flow hemodynamics.

 There have been numerous studies exploring MB localization based on Deep Learning (DL). 
Deep-ULM was introduced in \cite{Mischi_ICASSP2019}, to improve the precision and recall of standard localization methods, specifically in areas with high MB PSF overlaps while also maintaining low computational complexity.
Moreover, to improve MB localization in high-density regions, an encoder-decoder based Convolutional Neural Network (CNN)  with radiofreqyency (RF) and envelope data as the inputs was utilized in \cite{Song_TUFFC2022}.
In \cite{Milecki_TMI2021}, a 3D CNN was used for recovering super-resolution maps, considering spatiotemporal information. Since the ULM procedure was solved as a binary task to minimize Dice loss, MB tracking is not directly plausible using their method and hence velocity maps could not be generated.

 In addition to the above mentioned challenges, these studies failed to consider the underlying tissue signals, as they based their assumption on suppressing these signals using clutter filters such as Single Value Decomposition (SVD). This added a pre-processing step, adds to the manually adjusted parameters needed for MB localization, which in turn, brings forth errors that could possibly propagate throughout the ULM process. Furthermore, this questions the adaptability of these methods for different imaging settings.
 
 Transformers and attention modules \cite{Vaswani_attn2017} have been recieving a wide attention in image processing fields. A self-attention module enables us to have larger receptive fields and avoid CNNs' inductive bias.
Since transformers are trained on large datasets, fine-tuning them allows us to take advantage of large expert-annotated datasets and has resulted in valuable solutions for many tasks in imaging. In recent years, there has been an increasing interest in transfer learning using transformers in medical imaging \cite{Gheflati _EMBC2022}.

 Previously, we studied the potential of transformer-based solutions for MB localization with DEtection TRansformer (DETR) \cite{Gharamaleki_IUS2022}. We achieved high precision and recall in different imaging settings, as one of the finalists to the IEEE IUS UltraSR challenge. To improve on our previous work, we investigate the use of DEformable DETR (DE-DETR) \cite{Xie_DEDETR_CVPR2021}, to increase the MB localization accuracy, while decreasing the computational cost of training and testing. 
We validate the results of our network on the simulation dataset and compare the results against DETR and the ground truth.

\section{Methods}
To further improve the efficiency of our proposed network in \cite{Gharamaleki_IUS2022}, we have pre-trained the fine-tuned DE-DETR network. The simulation dataset along with the ground truth are provided by the IEEE UltraSR challenge organizers.

 The simulation video frames were masked using the ground truth MB locations to present all MB bounding boxes along with their centroids. Thereafter, individual MB annotations were extracted to COCO (Common Objects in COntext) \cite{Microsoft_ECCV2014} format to form the input for the DETR and DE-DETR networks.
\subsection{DETR}
\begin{figure*}[b]
  \centering

    \subfigure[]{\includegraphics[width=0.2\textwidth,height=0.4\textwidth]{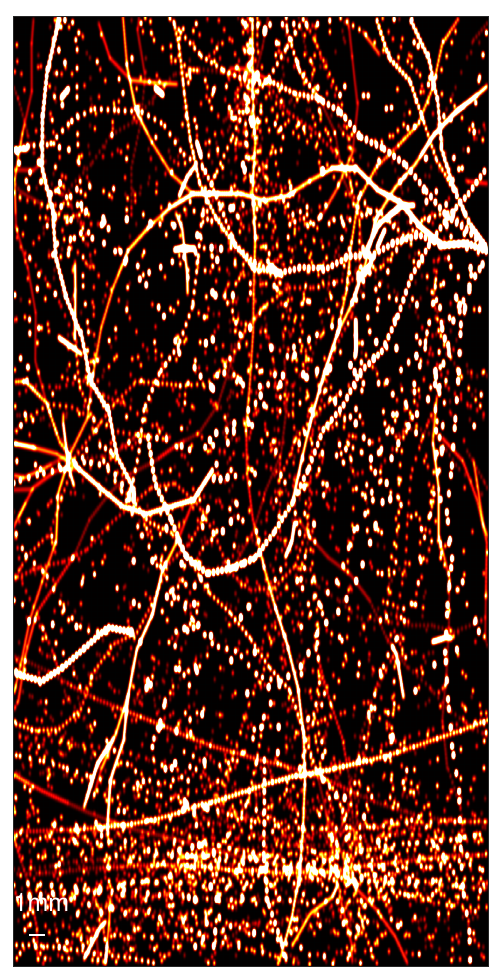}} 
    \subfigure[]{\includegraphics[width=0.2\textwidth,height=0.4\textwidth]{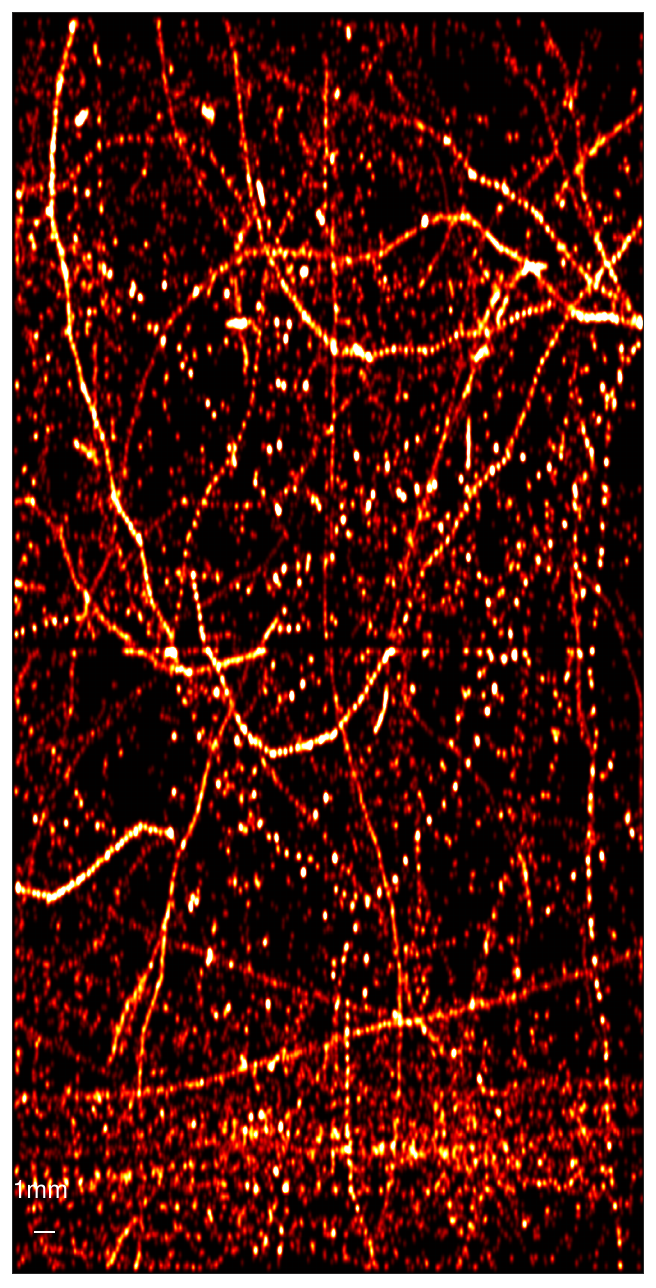}}
    \subfigure[]{\includegraphics[width=0.2\textwidth,height=0.4\textwidth]{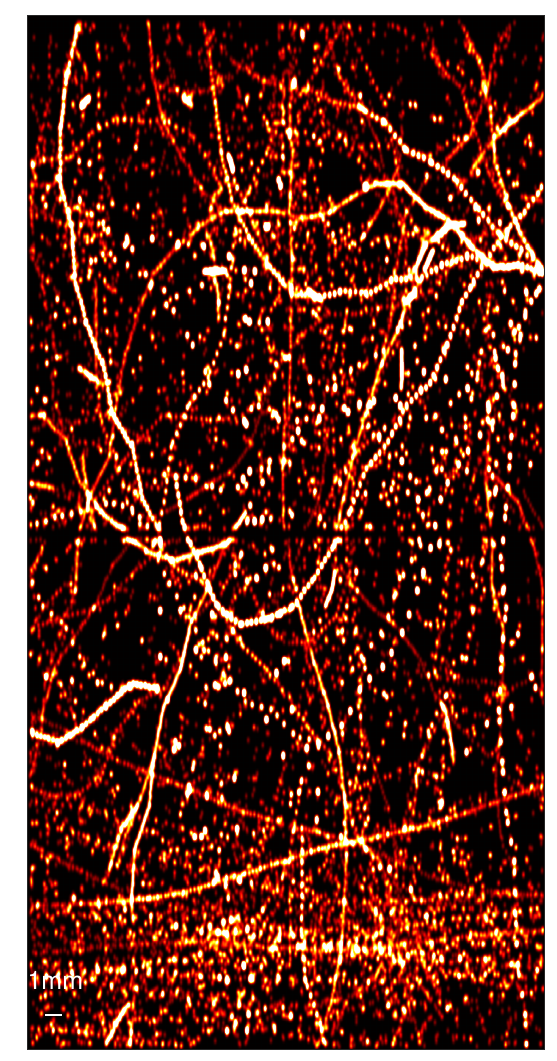}}

  \caption{SR images of the test dataset, i.e., the last 100 frames of a simulation video from the challenge dataset obtained from (a) Ground Truth (b) DETR and (c) DE-DETR.}
  \label{fig:simucompare}
\end{figure*}

DETR's architecture as shown in Figure \ref{fig:DETR} consists of a CNN backbone (ResNet50 \cite{Resnet_CVPR2016}), which extracts image features, encoders to interpret image information and decoders to generate the predictions.
The encoder-decoder structure consists of multi-head attention and self-attention, which allows the network to use the entire image as the context by taking into account their pair-wise correlations. Using this method, the network can decide on the class and bounding boxes of all the objects globally. 

A multi-layer preceptron is also used as prediction head to map the output of decoder to a bounding box and a class.   
Feature map pixels with added positional embeddings are the queries to the encoder and learnt object queries, as defined in \cite{Carion_DETR_ECCV2020}, are the decoder queries.

 To avoid near-duplicate predictions and enhance permutation-invariance, a set prediction loss based on bipartite matching is utilized to find optimal matches. Using bipartite matching and by considering the background as a "no object" class, each object is ensured to have a unique bounding box. Since a set prediction loss is deployed to find the final predictions, object queries need to be more than the maximum number of objects in all the frames of video.

 After finding the best matches using bipartite matching, a loss function which is a linear combination between a cross entropy loss and the bounding box error is defined using the Hungarian algorithm, the details of which is mentioned in \cite{Carion_DETR_ECCV2020},

\subsection{DE-DETR}
A number of challenges remain to be addressed when using DETR for MB localization.
The quadratic computational cost of the encoder, as a result of casting attention to all of the input grid pixels for each input reference point, leads to a slow convergence for DETR. 

Furthermore, as shown in the original paper \cite{Carion_DETR_ECCV2020}, the network does not perform as well for smaller objects due to its limited feature resolution.

 DE-DETR \cite{Xie_DEDETR_CVPR2021} was introduced to improve DETR's low scalability by incorprating a multi-scale deformable attention mechanism, as visualized in Figure \ref{fig:DEDETR}.

Distributing attention more thoughtfully and working with a limited attention budget is what distinguishes DE-DETR from DETR, which uses attention on every pixel of the input.
Furthermore, multiple scales of feature map is used, as opposed to DETR, to increase the localization accuracy for smaller objects. 
For every reference point on the input feature map, we calculate a set of offsets, using the queries, which guide where attention should be directed. The details of calculating offsets are outlined in \cite{Xie_DEDETR_CVPR2021}.
By ensuring that the number of sample keys for offsets remains smaller than the total number of pixels in the  image, the network's complexity maintains a non-quadratic trajectory.

\subsection{Fine-tuning}
Adopting DL techniques in medical imaging has relied heavily on transfer learning and fine-tuning. By building on top of the DL solutions for similar vision tasks, medical imaging tasks might benefit from knowledge learnt beforehand. Having in mind that transformers are particularly "data-hungry", this prior knowledge can contribute significantly to the network's ability to have higher precision in scenarios where annotated data are scarce and to decrease the computational complexity. 

Despite this, there are fundamental differences between common vision images and medical images, such as data size and features.  Using the dataset from the challenge and the pre-trained weights, we have fine-tuned the model to account for the significant domain shift between the two areas.
\section{Results}
 DETR's training settings are set as our previous implementation from \cite{Gharamaleki_IUS2022}.  The pre-trained DE-DETR was fine-tuned on a subset of frames provided by the challenge. This subset comprised 100 frames, selected based on their low correlation ($<$18\%) with the remaining frames as the test dataset (to prevent data leakage from training to test). The rest of the frames were used for training the network. 
 
 We evaluate the accuracy of identifying individual MBs in the evaluation. Mean Average Precision (mAP) and mean Average Recall (mAR) are calculated for each network. 
As mentioned in \cite{Carion_DETR_ECCV2020} and \cite{Gharamaleki_IUS2022}, the number of object queries should always be considerably higher than the number of objects in each and all of the images. For this reason, the images are divided into four patches each to refrain from increasing the number of object queries and hence the computations. Additionally, to mitigate duplicate localizations of MBs along patch borders,  an algorithm was developed. Additionally, to enhance the generalization capability of the network, we utilized random image augmentations for both networks.

 The mAP and mAR for DETR are: 80.12\% and 55.17\% and for DE-DETR they are: 87.60\% and 63.79\%. The super-resolution maps are generated using a proper Gaussian around each MB location and shown for comparison in Figure \ref{fig:simucompare}.

\section{Conclusions}
In this study, we leveraged DE-DETR, a network that introduces a deformable attention mechanism to improve the accuracy of object detection. Deformable attention enables the model to better handle object deformations and variations in appearance, resulting in improved detection performance over DETR. This is crucial for MB localization, given the variability in MB size and appearances over time. Our method achieves super-resolution ultrasound without any pre or post processing algorithms, by adopting transfer learning. This also reduced our time and hardware dependencies.
We added the benefits of random augmentation of the dataset to introduce more generalization to the network. Utilizing DE-DETR was shown to increase the efficiency compared to DETR both by reducing the computational complexity of training and the accuracy for smaller objects detection.
Our next phase involves conducting phantom imaging experiments using MBs to validate and  refine our network.
Extending the results of our network to encompass in vivo images is another project we’re currently working on.

\section*{Acknowledgment}
We acknowledge the support of the Natural Sciences and Engineering Research Council of Canada (NSERC). B.H. holds a Burroughs Wellcome Fund CASI award

\end{document}